\documentclass[12pt]{article}

\usepackage{amsmath}
\usepackage{epsfig}
\usepackage{subfigure}

\def\e{\begin{equation}}
\def\f{\end{equation}}
\def\*{^{\displaystyle*}}
\def\%#1{\mbox{\boldmath $#1$}}
\def\_#1{{\bf #1\mit}}

\def\=#1{\overline{\overline #1}}

\def\M{\mu}

\def\.{\cdot}
\def\x{\times}
\def\##1{{\bf#1\mit}}

\def\l#1{\label{eq:#1}}
\def\r#1{(\ref{eq:#1})}
\def\am{\left(\begin{array}{c}}
\def\amm{\left(\begin{array}{cc}}
\def\a{\end{array}\right)}

\begin{document}

\title{Evanescent modes stored in cavity resonators with backward-wave slabs}
\author{S.A. Tretyakov, S.I. Maslovski, I.S. Nefedov, M.K. K\"arkk\"ainen
\\
Radio Laboratory, Helsinki University of Technology\\
P.O. Box 3000, FIN-02015 HUT, Finland\\
Tel: +358-9-4512243 \\
Fax: +358-9-4512152 \\
\\
E-mails: sergei.tretyakov@hut.fi, NefedovIS@info.sgu.ru, \\
stas@cc.hut.fi, mkk@cc.hut.fi}

%\date{Radio Laboratory, Helsinki University of
%Technology\\ P.O.\ Box 3000, FIN--02015 HUT, Finland\\ \today}

\maketitle

\begin{abstract}

As was shown by N. Engheta,
electromagnetic fields in two adjacent slabs
bounded by two metal walls can satisfy the
boundary conditions even if the distance between the
two walls is much smaller than the wavelength. This is possible if
one of the slabs
has a negative permeability.
Here we show that these
subwavelength ``resonators"
resonate only if the permeability of at least one
of the slabs is frequency dependent.
Thus, there is no advantage of using these structures as
frequency-selective devices. However, we have found that
these systems can be in principle used as memory devices for
evanescent fields,
because the boundary conditions are identically satisfied for
all plane evanescent waves inside the cavity.
The physical meaning and practical limitations for
possible realizations are discussed. The analysis is supported
by FDTD simulations.

\end{abstract}

\medskip
\noindent {\bf Key terms:} evanescent modes, subwavelength
resonator, Veselago materials, left-handed materials, double
negative materials, FDTD
\medskip

\newpage

%\vskip 2cm

%\newpage

\section{Introduction}

Recently, a lot of attention has been payed to
electromagnetic properties of materials with negative
parameters (called Veselago media, backward-wave media, double negative materials,
left-handed materials). Conceptualized by V.G. Veselago \cite{Veselago}
and realized by R.A. Shelby et. al. as a composite with metal inclusions of specific shapes
\cite{exp}, these materials are very much debated because of
their exotic properties and potential applications.
One of the applications is perfect lens, (theoretically) capable to
focus not only propagating, but also evanescent modes \cite{lens}.
Contradictory opinions about realizability of this device
have been published in the literature  \cite{contra}.

Another exciting idea is a possibility to design
subwavelength resonators \cite{Engheta}.
N.\ Engheta has shown that a pair of plane waves traveling
in the system of two planar slabs positioned between two metal planes
can satisfy the boundary conditions on the walls and on the
interface between two slabs even for arbitrarily thin
layers, provided that one of the slabs has negative material parameters.
In paper \cite{Engheta}, waves propagating in the direction
orthogonal to the interface and the metal walls
have been considered. The possibility to satisfy the
boundary conditions for small distances between metal plates
is based on the fact that plane waves in Veselago media are backward waves,
meaning that the phase shift due to propagation in
a usual slab can be compensated by a negative phase shift
inside a backward-wave slab.

\begin{figure}[h]
\centering
\epsfig{file=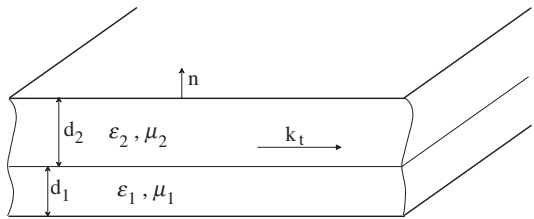, width=7cm}
\caption{Planar two-layer waveguide. One of the slabs can be a Veselago medium.}
\label{5veselo-two}
\end{figure}

In this paper, we analyze the system proposed by Engheta,
both analytically and numerically, for arbitrary wave solutions.
The geometry of the problem is shown in Figure~\ref{5veselo-two}:
it is a two-layer planar waveguide, and we study general
field solutions in form of propagating or evanescent modes with
arbitrary tangential propagation factors $\_k_t$.
The results allow better understanding of the phenomenon
predicted by N.\ Engheta, and lead to important conclusions regarding
evanescent modes in  the system. It will be shown that
evanescent excitations can be ``stored" in the resonator
volume after the external field has been removed.

\section{Eigenvalue equations}

Let us assume that the field between the two metal walls with  an arbitrary distribution in a plane orthogonal to $\_n$ is
expanded into a Fourier integral or series in that plane.
The field harmonics are plane waves traveling in the transverse
plane with two-dimensional wave vectors $\_k_t$.
The corresponding propagation factors along $\_n$ we denote by
\e \beta_{1,2}=\sqrt{k_{1,2}^2-k_t^2}\l{betas} \f
In the Fourier domain,
the exact boundary condition on the free interface of
a slab backed by an ideally conducting surface reads (e.g., \cite{vtc}):
\e \_E_{t+}=j\omega \M_{1,2}  {\tan{\beta_{1,2} d_{1,2}} \over{\beta_{1,2} }}\,
\=A_{1,2}\. \_n\x\_H_{t+}
 \f
 where
 \e \=A_{1,2}=\=I_t-{\_k_t\_k_t\over{k_{1,2}^2}}={\beta_{1,2}^2\over{k_{1,2}^2}} {\_k_t\_k_t\over{k_t^2}}   +
 {\_n\x\_k_t\, \_n\x\_k_t\over{k_t^2}}
\l{A}\f
 Because the
tangential fields $\_E_{t+}$ and $\_n\x\_H_{t+}$ are continuous on the interface
between the two slabs, we can write
\e\left(j\omega \M_1  {\tan{\beta_1 d_1} \over{\beta_1 }}\=A_1
+j\omega \M_2  {\tan{\beta_2 d_2} \over{\beta_2}}\=A_2\right)
\.\_n\x\_H_{t+}=0\l{5two-la}\f

Solution for the eigenwaves is now very easy because dyadics
$\=A_{1,2}$ are diagonal with the same set of eigenvectors: $\_k_t$ and $\_n\x\_k_t$.
Writing the two-dimensional vector
$\_n\x\_H_{t+}$ in this basis:
\e \_n\x\_H_{t+}= a\, \_k_t/|k_t|+b\, \_n\x\_k_t/|k_t|\f
and substituting into \r{5two-la}, we arrive to
equations for the propagation constant $k_t$.
If $a\neq 0$ and $b=0$, vector $\_H_{t+}$ is directed along
$\_n\x\_k_t$, that is, orthogonal to the propagation direction. This gives
the TM mode solution. The eigenvalue equation in this case is
\e {\beta_1\over{\epsilon_1}}\tan{\beta_1 d_1} +{\beta_2\over{\epsilon_2}}\tan{\beta_2 d_2}=0
\l{5TM}\f
For the other mode, when $b\neq 0$ and $a=0$,  the magnetic field vector is along
$\_k_t$ (TE mode), and we get
\e {\mu_1\over{\beta_1}}\tan{\beta_1 d_1} +{\mu_2\over{\beta_2}}\tan{\beta_2 d_2}=0
\l{5TE}\f
Note that the square root branch defining the normal components of
the propagation factors \r{betas} can be chosen arbitrarily, which is
natural for the system with standing waves.

Let us now assume that the thicknesses of the two layers are the same
($d_1=d_2=d$), and the material parameters differ by sign: $\mu_2=-\mu_1=-\mu$, $\epsilon_2=-\epsilon_1=-\epsilon$. In this particular case
the eigenvalue equations \r{5TM} and \r{5TE} are satisfied
{\it identically for all $\_k_t$}! This result can be perhaps
better understood deriving the matrix relation between tangential fields
on the two metal boundaries.
The exact solution for tangential fields on the two sides of an isotropic  slab
can be conveniently written in matrix form as (e.g. \cite{vtc}):
\e \am \_E_{t+} \\ \_n\x\_H_{t+} \a
=\amm \=a_{11} & \=a_{12} \\ \=a_{21} & \=a_{22} \a
\. \am \_E_{t-} \\ \_n\x\_H_{t-} \a  \l{aaa} \f
The dyadic coefficients in this matrix read
\e \=a_{11}=\=a_{22}=\cos(\beta d)\=I_t \l{5a11_a22}\f
\e \=a_{12}={j\omega \mu \over\beta}\sin(\beta d)\=A,\qquad \=a_{21}={j\omega \epsilon\over{\beta}}
\sin(\beta d)\=C \l{5_a12_a21}  \f
where $\=I_t$ is the transverse unit dyadic,
$\=A$ is given by \r{A}, and
\e \=C=\=I_t- {\_n\x\_k_t\, \_n\x\_k_t\over{k^2}} =  {\_k_t\_k_t\over{k_t^2}}   +
{\beta^2\over{k^2}} {\_n\x\_k_t\, \_n\x\_k_t\over{k_t^2}}\l{1dyadicC}
\f
The total transmission matrix for the system of two slabs is the product of
matrices \r{aaa} for individual slabs.
For the
slab with negative parameters we have
\e \am \_E_{t+} \\ \_n\x\_H_{t+} \a
=\amm \=a_{11} & -\=a_{12} \\ -\=a_{21} & \=a_{22} \a
\. \am \_E_{t-} \\ \_n\x\_H_{t-} \a   \f
Simple calculation\footnote{Calculating the product, note that
$\overline{\overline{A}}\cdot\overline{\overline{C}}={\beta^2\over{k^2}}
\overline{\overline{I}}_t$. }
  shows that in this particular case
\e \amm \=a_{11} & \=a_{12} \\ \=a_{21} & \=a_{22} \a \cdot
\amm \=a_{11} & -\=a_{12} \\ -\=a_{21} & \=a_{22} \a
=   \amm  1 & 0 \\ 0 & 1 \a \cdot\=I_t \f
{\it identically for any} $\_k_t$ and {\it arbitrary thickness $d$ of the layers}.
This fact means that the boundary conditions at the metal walls
are satisfied not only for normally propagating waves, but also  for
obliquely traveling waves, and, most important, for {\it all evanescent modes}.

\section{Discussion}

Let us first discuss the possible use of this system as a
subwavelength resonator, as proposed by N.~Engheta \cite{Engheta}.
If the propagation constant along the slabs $\_k_t=0$, both
equations for TM and TE modes reduce to
\e {\mu_1\over{k_1}}\tan{k_1 d_1} +{\mu_2\over{k_2}}\tan{k_2 d_2}=0
\l{5reson}\f
that is the resonance condition for standing waves in the dual-layer system between two
metal plates. If the thicknesses of both layers are small compared with the
wavelength, one can simplify this equation replacing tangent functions by the
first terms of their Taylor expansions:
\e \mu_1d_1+\mu_2d_2=0\l{5thin-res}\f
From here it is obvious that
if both  permeabilities are positive (or both negative), no resonance is possible in thin
layers: the thickness should be of the order of the wavelength (half-wavelength resonance).
However, as was noticed by N.\ Engheta \cite{Engheta},
if {\it one} of the permeabilities is negative, condition
\r{5thin-res} can be satisfied even for very thin layers. This appears to
open a possibility to realize very compact resonant cavities.

However, let us think what kind of resonator we get this way?
Resonance as such means that a certain circuit function (reflection coefficient,
input impedance\dots) sharply changes with the frequency. In the
conventional case of positive media parameters, the two-layer cavity is a
{\em resonator} because the tangent functions in \r{5reson}
quickly vary with respect to the frequency near the resonance. Now look at
relation \r{5thin-res}: there is no explicit dependence on the frequency at all!
So, if the material parameters are assumed to be approximately frequency independent over a
certain frequency range, the system does not resonate, although the
``resonance condition" \r{5thin-res} is satisfied. Indeed, equation \r{5thin-res} is
satisfied for all frequencies in this range, thus this system is not
frequency selective. What actually happens is that the inductive reactive part of the
input impedance of a usual layer ($+j\omega \mu_1d_1$) is compensated
by the {\em negative  inductive impedance} of the other layer
($-j\omega |\mu_2|d_2$). What is left is only resistance, defined by losses in the
materials of the slab.
Of course, in realistic situations material parameters are
frequency dependent, but since the frequency variations
in the response functions are determined by the permeability only
and do not depend on the frequency explicitly,
the resonant phenomena in subwavelength
resonators suggested by Engheta are determined only by the resonant properties of
the permeability function of the material layers. In this sense
phenomena in thin resonant layers resemble resonance  in reflection from
a thin ferrite layer on a metal plane or from a ferrite sphere near a
microstrip line or in a closed waveguide.

However, the most interesting case is when the waves
between the planes are evanescent,
meaning that $k_t^2>|k_{1,2}|$. In this case
the waves actually decay in the direction
orthogonal to the interface, because the corresponding propagation factors
$\beta_{1,2}=\sqrt{k_{1,2}^2-k_t^2}$ are imaginary.
Suppose that the volume between the two metal screens is excited by
a source with a certain fast variation of the current or field in space.
Under the above assumptions regarding the media properties, all evanescent modes of this source satisfy the boundary conditions. Conceptually, this is an ideal
memory device, because after the source has been removed, the field distribution
near the interface of the two slabs will be preserved (until losses will
consume the field energy), and the
field distribution will correspond to the field distribution of the source.

This unique property wholly depends on the assumption that in one of the
slab the material parameters are negative. The slab thickness is irrelevant, and
the same effect exists also on a single interface between two half spaces filled by the same materials as the two slabs in the ``resonator". Indeed, the ``memory"
effect for evanescent modes with arbitrary propagation constants is due to the
fact that the interface supports surface modes with arbitrary propagation
constants, at the frequency where $\mu_2=-\mu_1$ and $\epsilon_2=-\epsilon_1$.
Actually, the same phenomenon is the core effect that makes the planar
slab of a Veselago medium act as a perfect lens \cite{lens}. In that device,
there are two such interfaces supporting surface waves with arbitrary
wavenumbers. Resonant excitation of these modes leads to amplification of
evanescent waves crossing the slab.

The main assumption has been that the Veselago material is
an effective magnetodielectric {\it medium,} and the main challenge in
realizing any device using the principle explained here is to design
such a material with as small spatial period as possible.
Higher-order evanescent modes vary extremely fast in space, and
as soon as the spatial period of the exciting field becomes
comparable with the spatial period of the artificial material with
negative parameters, spatial dispersion effects  degrade
the properties of the device. Clearly, there are  other limitations
related to absorption present in any realistic medium and to the final size
of the slabs
in the transverse direction.

\section{FDTD simulated behavior}

\begin{figure}[h]
\centering
\epsfig{file=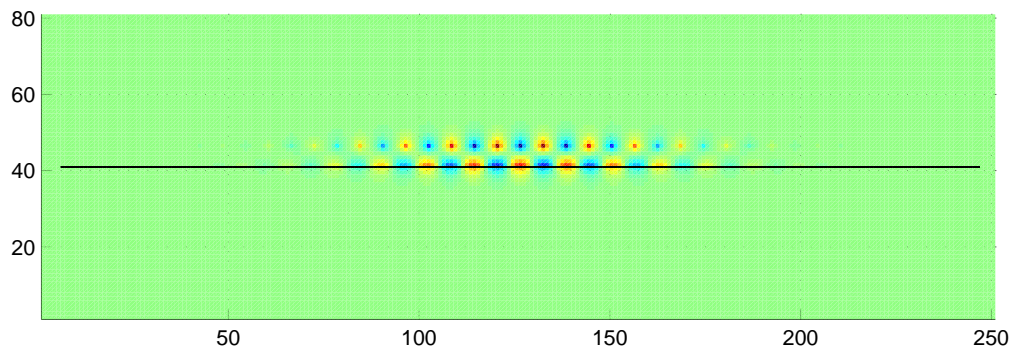, width=15cm}
\caption{Evanescent field distribution near an interface with
a backward-wave material region. The source (over the interface) excites an
eigenmode of the interface. Cell numbers shown near the grid edges.}
\label{eva1}
\end{figure}

\begin{figure}[h]
\centering
\epsfig{file=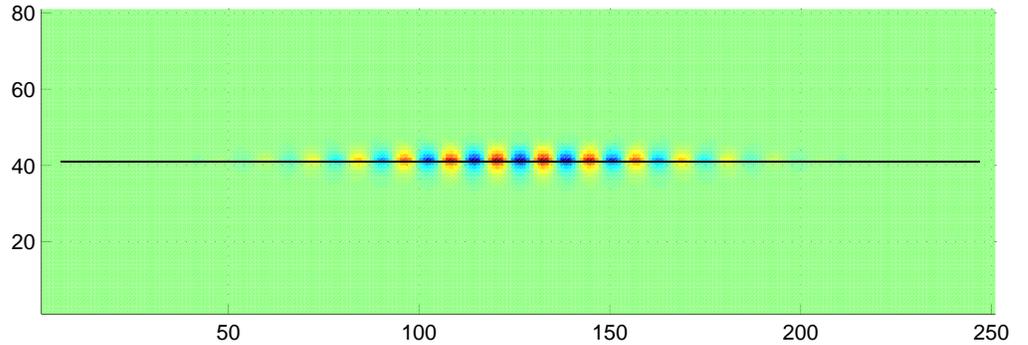, width=15cm}
\caption{The same as in Figure~\ref{eva1} at a later moment of time.
The source field has decayed to zero, but the field
excited near the interface remains and its spatial
distribution is very similar to that of the original source.}
\label{eva2}
\end{figure}

We have simulated fields near an interface between free space and
a backward-wave medium slab.
Negative permittivity and permeability are assumed to
follow the
Lorentz dispersion model:
\begin{equation}
\epsilon(\omega) = \epsilon_0 \left( 1+
\frac{\omega_{pe}^2}{\omega_{0e}^2-\omega^2+j\Gamma_e \omega}
\right), \quad \mu(\omega) = \mu_0 \left( 1+
\frac{\omega_{pm}^2}{\omega_{0m}^2-\omega^2+j\Gamma_m \omega}
\right) \label{eq:perms}
\end{equation}
This model corresponds to a realization of Veselago materials as
mixtures of conductive spirals or omega particles, as discussed in
\cite{Tretyakov}.
Used discrete FDTD model of the material is
based on the constitutive relation discretized after one
integration. This approach, described in detail in
\cite{Mikko},   leads to
much better accuracy and, what is also important, to considerably
better stability than the conventional direct discretization without
integration.
In the simulations, the loss factors have been set to zero, and
the other parameters chosen so that at the central
frequency of the source spectrum both relative parameters equal $-1$.

\begin{figure}[h]
\centering
\epsfig{file=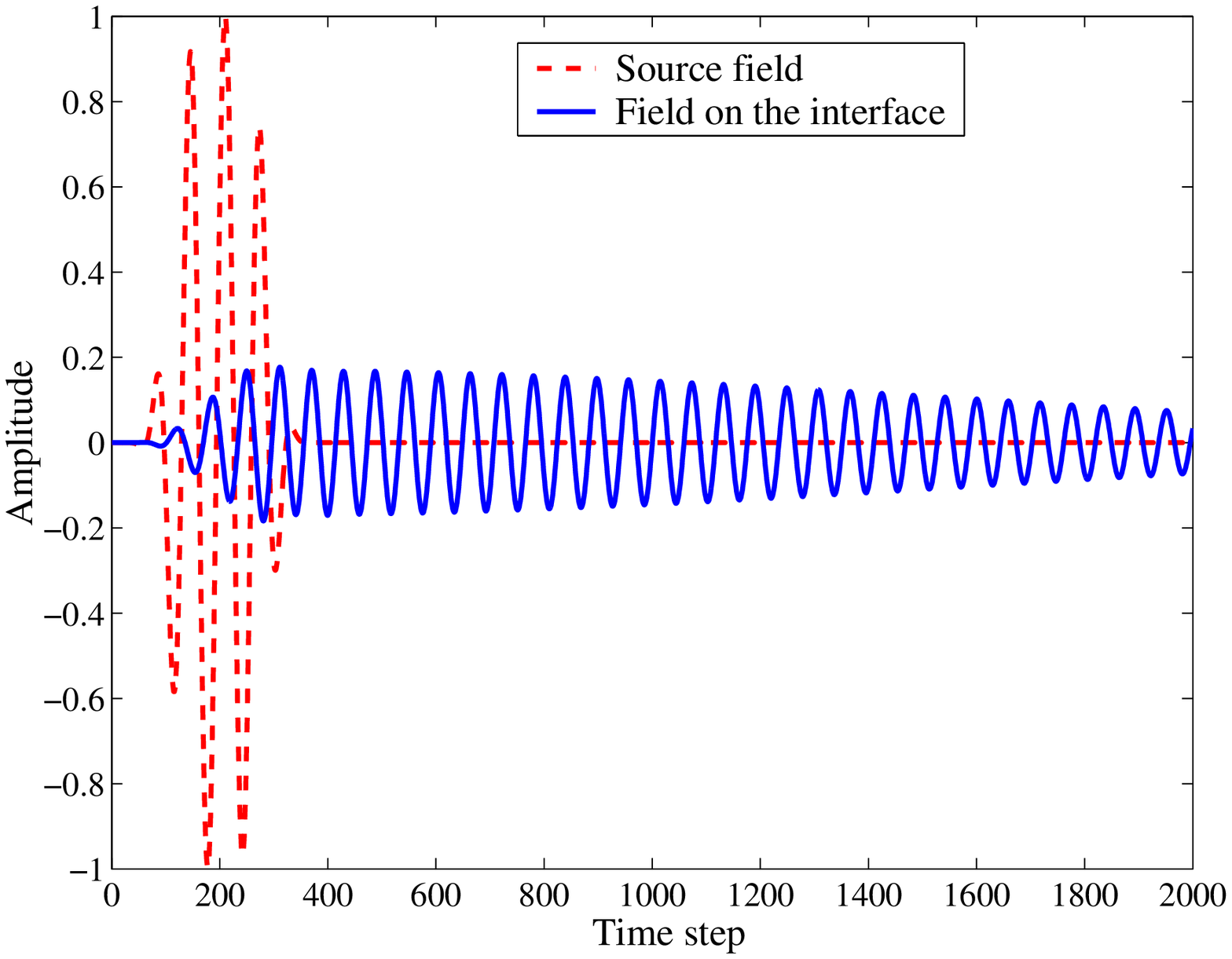, width=13cm}
\caption{Time dependence of the field of the source and on the
interface. It is seen that the field at the interface continues
oscillations with the same frequency as that of the source
after the source has been switched off.}
\label{time}
\end{figure}

To visualize the behavior of evanescent modes, the
interface has been excited by an array of line sources with
the period
smaller than the free space wavelength. Each line source is out-of-phase with
its two nearest sources. In the numerical example,
the source spectrum is concentrated near 0.4377 GHz, corresponding to 68.5 cm wavelength.
The distance between the line sources is 9 cm (6 cells, cell size is
1.5 cm). Thus, the incident field exponentially decays in the direction
orthogonal to the source plane and forms a standing wave with
a small period in the source plane. The distance from the source to the
interface is equal to the array period (9 cm).
The time dependence of the source is
\e E_{\rm inc}(t)= e^{-\left({t-t_0\over{t_1}}\right)^4}\sin(\omega_0 t) \f
where $t_0=200\Delta t$, $t_1=100 \Delta t$, and $\Delta t$ is the time step.

Computed spatial distributions of the field amplitude are
shown in Figures~\ref{eva1} and \ref{eva2}.
The time dependence of the field in the source plane and at the
plane of the interface is plotted in
Figure~\ref{time}. The simulated results confirm the
theoretical predictions.
First, the source excites oscillations near the
interface between free space and the backward-wave medium, because the
interface supports modes with arbitrary propagation
constants along the interface plane. Next, the  source is switched
off, but the oscillations near the boundary stay there and
get distorted  very slowly (due to medium dispersion), since
the field satisfies the Maxwell equations and the boundary conditions
at the interface. In this numerical example, electric walls are not
present, and it is obvious from the results that they are not
relevant for the memory effect for evanescent modes:
evanescent fields are concentrated only near the source and the
media interface.

\section{Conclusion}

Detailed analysis of field solutions in thin subwavelength
cavity resonators using materials with negative permittivity and
permeability reveals that these cavities can in principle
support evanescent fields concentrated near the interface of the two
slabs. Numerical simulations show that when the evanescent field source
is removed, resonant field excited near the interface continues to
oscillate for a long time, as expected from the analytical analysis.
Numerical experiments also indicate that the system is rather
sensitive to the material parameter values: slight deviations from the
resonant values reduce the excited field amplitude.
The analysis of the cavity as a resonator for waves traveling along $\_n$
shows that this system can be used as a frequency selective device
only due to resonant frequency dependence of the material parameters of
one or both materials slabs.

\end{document}